\documentclass[iop]{emulateapj}
\slugcomment{Accepted by {\it The Astrophysical Journal}}
\def\farcm{\hbox{$.\mkern-4mu^\prime$}}
\def\farcs{\hbox{$.\mkern-4mu^{\prime\prime}$}}

\def\la{\mathrel{\hbox{\rlap{\hbox{\lower4pt\hbox{$\sim$}}}\hbox{$<$}}}}
\def\ga{\mathrel{\hbox{\rlap{\hbox{\lower4pt\hbox{$\sim$}}}\hbox{$>$}}}}

\shortauthors{Park}
\shorttitle{N49B}

\begin{document}
\title{Spatial Distribution of Mg-Rich Ejecta in LMC Supernova Remnant N49B }

\author{Sangwook Park\altaffilmark{1} and Jayant Bhalerao}

\affil{Box 19059, Department of Physics, University of Texas at Arlington,
Arlington, TX 76019}

\altaffiltext{1}{s.park@uta.edu}

\begin{abstract}

The supernova remnant (SNR) N49B in the Large Magellanic Cloud is a peculiar 
example of a core-collapse SNR to show the shocked metal-rich ejecta enriched 
only in Mg without evidence for a similar overabundance in O and Ne. Based 
on archival {\it Chandra} data we present results from our extensive 
spatially-resolved spectral analysis of N49B. We find that the Mg-rich ejecta 
gas extends from the central regions of the SNR out to the southeastern 
outermost boundary of the SNR. This elongated feature shows an overabundance 
for Mg similar to that of the main ejecta region at the SNR center, and its 
electron temperature appears to be higher than the central main ejecta gas.
We estimate that the Mg mass in this southeastern elongated ejecta feature 
is $\sim$10\% of the total Mg ejecta mass. Our estimated lower limit of 
$>$0.1 $M_{\odot}$ on the total mass of the Mg-rich ejecta confirms the 
previously-suggested large mass for the progenitor star ($M$ $\ga$ 25 
$M_{\odot}$). We entertain scenarios of an SNR expanding into a non-uniform 
medium and an energetic jet-driven supernova in an attempt to interpret 
these results. However, with the current results, the origins of the extended 
Mg-rich ejecta and the {\it Mg-only-rich} nature of the overall metal-rich ejecta 
in this SNR remain elusive.

\end{abstract}


\section {\label {sec:intro} INTRODUCTION}

N49B is an X-ray-bright supernova remnant (SNR) in the Large Magellanic Cloud 
(LMC).  The presence of a nearby H {\small II} region \citep{chu88} and a 
molecular cloud \citep{cohe88} suggested a core-collapse supernova (SN) 
explosion of a massive star for the origin of N49B. The shock expansion in 
an interstellar cavity was suggested by the {\it ASCA} data \citep{hugh98}, 
which was supportive of its massive progenitor star. Based on {\it Chandra} 
data we detected a large variation in the surface brightness of the swept-up 
interstellar medium (ISM), which indicates that the SNR is expanding into an 
ambient medium with an order of magnitude density variation along the outer 
boundary of the SNR, generally consistent with the shock interaction with 
nearby molecular clouds \citep[ P03 hereafter]{park03b}. The {\it Chandra} 
data allowed us to detect the shocked metal-rich ejecta in N49B for the first 
time (P03).  The detected metal-rich ejecta gas is enriched only in Mg, and 
the estimated large mass for the Mg-rich ejecta gas firmly established a very 
massive progenitor, and thus a core-collapse origin for N49B (P03). The 
putative compact stellar remnant has not been detected, and a 3$\sigma$ upper 
limit on the central source luminosity ($L_{0.5-5 keV}$ $<$ 2 $\times$ 10$^{33}$ 
erg s$^{-1}$) was placed based on the {\it Chandra} data (P03).

It is notable that only two SNRs, including N49B, (out of $\sim$500 SNRs 
detected in the Galaxy and Magellanic Clouds) show this characteristic 
Mg-overabundance (while {\it no other} metal elements are similarly enhanced): 
the Galactic SNR G284.3--1.8 is the other member of this peculiar group 
\citep{will15}. The true nature and origin of the {\it only-Mg-rich} ejecta 
in these SNRs are unclear. It might have resulted from a peculiar core-collapse 
nucleosynthesis, or a particular thermal condition in the ejecta gas might be 
responsible for the observed Mg line enhancement. If the observed Mg-rich nature 
(without enhancements in O and Ne) represents the true metal abundance structure 
in these SNRs, it may challenge the standard core-collapse nucleosynthesis models 
in which an invariably larger amount of O and Ne (than Mg) is expected to be 
created \citep[e.g.,][]{thie96,nomo06}. N49B is particularly important to study 
this rare group of SNRs because it appears that the bulk of the shocked Mg-rich 
ejecta gas has been detected (whereas the Mg-rich ejecta in G284.3--1.8 is 
detected in a small part of the SNR). The detection of the {\it entire} Mg-rich 
ejecta in N49B allows us a unique opportunity to study the overall nature of
the ejecta, such as the total shocked ejecta mass and its spatial structure, 
in this peculiar SNR.

We here report the presence of an emission feature in N49B, extending from the
central ejecta regions out to the southeastern outer boundary of the SNR. This
elongated region is enriched only in Mg just like the bulk of the metal-rich
ejecta gas in the central region of the SNR. This extended Mg-rich emission 
feature was hinted at by the Mg line equivalent width (EW) image in the previous 
{\it Chandra} data analysis (P03), but was not discussed in P03. In this work we 
focus on the spectral and morphological properties of this particular feature. 
We present the data and data analysis in Sections~2 \& 3, respectively. In 
Section~4 we discuss the nature of the elongated Mg-rich ejecta. 

\section{\label{sec:data} DATA}

N49B was observed with the S3 chip of the Advanced CCD Imaging Spectrometer 
(ACIS) on board {\it Chandra} on 2001 September 15 as part of the Guaranteed 
Time Observation program (P03). We reprocessed these archival {\it Chandra} 
data (ObsID 1041) of N49B (using CIAO version 4.2 with CALDB version 4.3.0). 
We processed the raw event file following the standard data reduction methods 
which include the correction for the charge transfer inefficiency. We applied 
the standard data screening by event status and grade ({\it ASCA} grades 02346). 
As noted by P03 the overall light curve from the entire S3 chip shows a 
moderately enhanced particle background (by a factor of $\sim$2 above the 
average quiescent count rate) for the last $\sim$10\% of the exposure. We 
excluded this part of the data ($\sim$4-ks time interval) from our analysis.  
After the data reduction the total effective exposure is $\sim$30.2 ks.

\section{\label{sec:res} ANALYSIS \& RESULTS}

\subsection{\label{subsec:jet} Mg Enhancement in Southeastern Region}

The broadband ACIS image of N49B shows a nearly circular overall morphology 
with several relatively bright filaments located mostly around the outer 
boundary of the SNR (Figure~\ref{fig:fig1}a). The central regions of N49B, 
in which X-ray emission is dominated by the shocked Mg-rich ejecta (P03), 
are generally faint and diffuse without showing any well-organized emission 
feature (corresponding to the Mg-enhanced regions). The Mg-rich ejecta gas 
was identified aided by a Mg K$\alpha$ line (at $E$ $\sim$ 1.35 keV) EW 
image (P03). In Figure~\ref{fig:fig1}b we reproduce the Mg line EW image 
of N49B as was presented in P03. To create Figure~\ref{fig:fig1}b we 
followed the same methods used by P03 (and references therein). We used 
the same photon energy bands for the Mg K$\alpha$ line ($E$ = 1280 -- 1440 
eV), and for the adjacent continuum ($E$ = 1140 -- 1240 and 1550 -- 1700 eV, 
to estimate the underlying continuum flux) as those selected by P03. While 
the bright Mg line EW in the central region of the SNR is evident, the 
enhanced Mg line clearly extends to the southeast (region E in 
Figure~\ref{fig:fig1}b). This enhanced line EW feature is elongated 
($\sim$20$^{\prime\prime}$ in thickness and $\sim$65$^{\prime\prime}$ 
in length), extending from the central Mg-rich ejecta region out to the 
outermost southeastern boundary of the SNR. This feature was hinted at 
in Figure~2 of P03 (although it appeared to be less clear than that in 
Figure~\ref{fig:fig1}b, probably due to a different smoothing method 
adopted by P03), but it was not discussed by P03 as they focused on the 
detection of the central Mg-rich ejecta. 

To quantitatively assess the apparently Mg-enhanced nature of this region 
we extracted the observed X-ray spectrum from region E (as shown in 
Figure~\ref{fig:fig1}). The X-ray emission in this region contains 
$\sim$7300 counts in the 0.4 -- 5 keV band. We binned the extracted 
spectrum to contain at least 20 counts per photon energy channel. We fit 
this spectrum with a non-equilibrium ionization (NEI) plane shock model 
(vpshock in XSPEC, Borkowski et al. 2001). We used NEI version 2.0 in 
XSPEC, associated with ATOMDB \citep{fost12} which includes inner-shell 
lines (from Li-like ions) and updated Fe-L shell transition lines 
\citep[see][]{bade06}. We note that the plane shock model used by P03 did 
not include these additional X-ray lines which could affect the estimates 
of the thermal parameters (and thus elemental abundances) for an 
under-ionized gas. In these spectral model fits we assumed two components 
for the foreground absorbing column: one for the Galactic absorption toward 
the LMC ($N_{H,Gal}$) and the other for the LMC column ($N_{H,LMC}$). We 
fixed $N_{H,Gal}$ at 6 $\times$ 10$^{20}$ cm$^{-2}$ \citep{dick90} with 
solar abundances \citep{ande89}. We fitted $N_{H,LMC}$ assuming the LMC 
abundances \citep{russ92,sche16}. 

For the plane shock component we initially fixed all elemental abundances 
at the LMC values: i.e., He = 0.89, C = 0.303, N = 0.123, Ar = 0.537, Ca 
= 0.339, Ni = 0.618 \citep{russ92}, and O = 0.13, Ne = 0.20, Mg = 0.20, 
Si = 0.28, S = 0.31, Fe = 0.15 \citep{sche16}. Abundances are with respect 
to solar values \citep{ande89}, hereafter. We fixed the redshift parameter 
at $z$ = 9.54 $\times$ 10$^{-4}$ for the radial velocity of the LMC 
\citep{meab95}. The electron temperature ($kT$, where $k$ is the Boltzmann 
constant) and ionization timescale ($\tau$ = $n_et$, where $n_e$ is the 
electron density and $t$ is the time since the gas has been shocked) are 
varied. The normalization parameter (a scaled volume emission measure, EM 
= $n_e$$n_H$$V$, where $n_H$ is the H density and $V$ is the emission volume) 
is also varied. The fit is statistically unacceptable ($\chi^2_{\nu}$ $\sim$ 
2.4) due to large residuals around the Mg K line feature ($E$ $\sim$ 1.35 keV). 
Also, there is significant excess emission at $E$ $\ga$ 2 keV above the 
best-fit model ($kT$ $\sim$ 0.6 keV). The observed X-ray spectrum in this 
high energy tail shows negligible (or very weak) K-shell line emission 
features from S ($E$ $\sim$ 2.4 -- 2.6 keV), Ar ($E$ $\sim$ 3.1 -- 3.3 keV), 
and Ca ($E$ $\sim$ 3.9 -- 4.1 keV). 

Varying the Mg abundance significantly improves the fit ($\chi^2_{\nu}$ 
$\sim$ 1.4), and the best-fit Mg abundance is $\sim$0.52, higher than 
the LMC value by a factor of $\sim$2.6, suggesting the presence of 
overabundant Mg. When we additionally varied abundances for O, Ne, Si, 
and Fe (whose K- and/or L-shell transition lines are prominent in our 
fitted 0.4 -- 5.0 keV band), the statistical improvement of the model 
fit is not significant ($\chi^2_{\nu}$ $\sim$ 1.3). The best-fit Mg 
abundance does not change (Mg = 0.52), while all other fitted abundances 
are consistent with the LMC values within uncertainties. It is notable 
that, even with varying abundances, the significant excess emission at 
$E$ $\ga$ 2 keV is evident for the best-fit electron temperature $kT$ 
$\sim$ 0.6 keV. Thus, although the best-fit one-temperature model may 
be considered to be statistically acceptable ($\chi^2_{\nu}$ $\sim$ 
1.3 -- 1.4), we added another NEI plane shock component in our spectral 
model to properly fit the excess emission at $E$ $>$ 2 keV. For the 
two-component model fit we initially fixed all metal abundances at the 
LMC values for both components, while varying other model parameters. 
The fit is statistically unacceptable ($\chi^2_{\nu}$ $\sim$ 2.0). It 
is clear that the best-fit model underestimates the strong Mg line at 
$E$ $\sim$ 1.35 keV. All other emission line features are adequately 
fitted. The bulk of X-ray emission at $E$ $\la$ 1 keV is fitted by the 
K lines from H- and He-like O and Ne ions (and some L-shell transition 
lines from Fe ions) with the LMC-like abundances. Thus, we fixed all 
metal abundances at the LMC values for the soft component assuming that 
this component is dominated by the emission from the shocked LMC ISM. 
We consider the hard component spectrum to originate primarily from the 
shocked Mg-rich ejecta, and thus varied the Mg abundance (with all other 
elements being fixed at the LMC abundances) for the hard component. The 
model fit significantly improves ($\chi^2_{\nu}$ $\sim$ 1.0). The 
best-fit Mg abundance for the hard component ($kT_{hard}$ $\sim$ 1.55 
keV) is 0.93. These results are summarized in Table~\ref{tbl:tab1}. The 
observed spectrum of region E (with this best-fit two-component plane 
shock model overlaid) is shown in Figure~\ref{fig:fig2}.  
In these spectral model fits of the region E spectrum we assumed 
the recent measurements of the LMC abundances for O, Ne, Mg, Si, and Fe 
(except for the Mg abundance in the ejecta component spectrum) based on 
the {\textit {Chandra}} data of the LMC SNRs \citep{sche16}, while 
we adopted LMC abundance values by Russell \& Dopita (1992) for other 
elements. We note that, for the LMC abundances, several measurements are available
in the literature \citep[e.g.,][]{russ92,hugh98,davi15,magg16,sche16}. 
There are some inconsistencies (up to by a factor of $\sim$2) among
those abundance measurements, whose origins are not fully understood: e.g., 
the cross-calibrations over different wavelengths and instruments, the 
use of different objects (stars, SNRs, H{\small II} regions), the model 
dependence of measurements, etc. We find no considerable effects on our 
spectral model fits depending on the choice of a specific LMC 
abundance set. The fitted parameters are consistent within
statistical uncertainties with equally-good fits: e.g., assuming the 
higher (and more traditional) values of the LMC abundances (O = 0.26, Ne = 
0.33, Si = 0.31, Mg = 0.32 and Fe = 0.36) measured by Russell \& Dopita (1992), we find
that the Mg abundance for the hard (ejecta) component is 1.12$^{+0.12}_{-0.11}$
with a nearly the same electron temperature ($kT$ $\sim$ 1.54 keV), 
ionization timescale ($\tau$ $\sim$ 0.7 $\times$ 10$^{11}$ cm$^{-3}$ s),
and the volume emission measure ($\sim$0.5 $\times$ 10$^{58}$ cm$^{-3}$) as
listed in Table~\ref{tbl:tab1}. 
For all other regional spectral analysis as presented in 
Section~\ref{subsec:other}, we fit the ISM abundances for O, Ne, Mg, 
Si, and Fe in our spectral model fits, while fixing other elemental 
abundances at values by Russell \& Dopita (1992). 

When we varied abundances for O, Ne, Si, and Fe (as well as Mg) in the 
hard component plane shock model fit, the statistical goodness-of-the-fit 
is nearly the same ($\chi^2_{\nu}$ $\sim$ 1.0) with similar main results: 
i.e., Mg is overabundant ($\sim$1.5) with a high electron temperature ($kT$ 
$\sim$ 4 keV). The Ne abundance appears to be elevated ($\sim$0.7) with a 
negligible Fe abundance ($<$0.04). These Ne and Fe abundances are somewhat 
different than those presented in Table~\ref{tbl:tab1}, and are potentially 
interesting to reveal the true nature of the ejecta composition and the SN 
nucleosynthesis of this SNR. However, the statistical improvement of this 
model fit is negligible, compared to that with the LMC abundances for those 
elements. Nonetheless, these (possibly non-LMC like) Ne and Fe abundances 
do not affect our conclusions on the Mg-rich ejecta in region E. Finally, 
we considered a power law (PL) spectrum to fit the high energy tail of the 
region E spectrum. The model fit (plane shock + PL) is statistically good 
($\chi^2_{\nu}$ $\sim$ 0.9). The Mg overabundance is evident ($\sim$0.92) 
with LMC-like abundances for all other fitted elements, but a low temperature 
($kT$ $\sim$ 0.34 keV) is implied for the Mg-rich ejecta gas. The best-fit PL 
photon index ($\Gamma$ $\sim$ 2.4) is consistent with that for the synchrotron 
radiation from relativistically-accelerated shocked electrons in the shell-type 
SNR N49B \citep{dick98}. However, the radio data of N49B show the {\it lack} 
of such non-thermal emission generally in the southern regions of the SNR 
\citep{dick98}, particularly in regions corresponding to region E (see 
Section~4). We consider that the physical origin of this 
phenomenologically-motivated PL model component is uncertain, and that the 
main results of the Mg-rich ejecta for region E are persistent. In the 
following discussion of region E, we assume the best-fit parameters from our 
two-component plane shock model fit with the LMC abundances for metal elements 
other than Mg (as presented in Table~\ref{tbl:tab1}), unless noted otherwise.

\subsection{\label{subsec:other} Central Ejecta and Ambient Gas}

We performed a similar spectral analysis of central Mg-rich ejecta regions. 
To characterize thermal condition and metal abundance for the central ejecta 
region, we extracted X-ray spectra from regions A -- D as marked in 
Figure~\ref{fig:fig1}. We selected these regions to represent {\it all} 
characteristic subregions showing the strong Mg line EW, while achieving
significant photon count statistics of $\ga$5000 counts per region. We fit 
these regional spectra with the NEI plane shock model generally following 
the method described in Section~\ref{subsec:jet}. As expected these regional 
spectra cannot be fitted by the LMC abundances ($\chi^2_{\nu}$ $\sim$ 3 -- 5) 
because of the significant discrepancy between the spectral model and the 
observed spectrum for the strong Mg K emission line feature. Varying the Mg 
abundance results in statistically acceptable fits ($\chi^2_{\nu}$ $\sim$ 
0.9 -- 1.3) for all four regional spectra. We note that, in contrast to region 
E, there is no excess emission above the best-fit model at $E$ $\ga$ 2 keV
for these central regions. The best-fit Mg abundance is significantly higher 
than the LMC value (by a factor of $\sim$4--6) for all four regions. While 
the LMC abundances for O, Ne, Si, and Fe can adequately fit the observed 
spectra of these central regions, for the purposes of statistical comparisons 
of metal abundances among the central ejecta regions and the shocked ISM 
regions, we fitted abundances for O, Ne, Si, and Fe as well as Mg in this 
work. As perhaps expected, best-fit abundances for O, Ne, Si, and Fe are 
consistent with the LMC values within statistical uncertainties, except for 
the moderately elevated Si abundance in region A (by a factor of $\sim$2 
above the LMC abundance). We summarize these results in Tables~\ref{tbl:tab2} 
\& \ref{tbl:tab3}. In Figure~\ref{fig:fig3} we show the observed spectra 
(with the best-fit NEI plane shock model overlaid) for regions A -- D.
  
We extract spectra from thirteen regions (regions 1 -- 13 in 
Figure~\ref{fig:fig1}) for which our Mg EW image shows no enhancements. 
X-ray emission in these regions presumably originates from the shocked ISM 
with LMC-like abundances (P03). We selected these regions to 
characteristically represent bright filaments and faint diffuse regions 
(in the broadband image) {\it exterior} to the enhanced Mg EW regions, 
while achieving significant photon statistics of $\ga$4000 counts in the 
0.4 -- 5 keV band. The observed X-ray spectra of these regions show weak 
atomic emission line features for all metal elements in contrast to regions 
A -- E in which a strong Mg line emission feature is evident. We show four 
example regional spectra from these shocked ISM-dominated regions in 
Figure~\ref{fig:fig4}. We performed the NEI plane shock model fits to 
these regional spectra with all metal abundances fixed at the LMC values.
These model fits are acceptable with $\chi^2_{\nu}$ $\sim$ 0.8 -- 1.4.
When we varied metal abundances for O, Ne, Mg, Si, and Fe, fits are nearly
the same ($\chi^2_{\nu}$ $\sim$ 0.8 -- 1.3) with similar best-fit values for
the fitted parameters. Although these regional spectra can be fitted with 
metal abundances fixed at the LMC values, for the purposes of statistical 
comparisons with the ejecta-dominated regions, we use the results based on 
our spectral model fits with elemental abundances varied in this work. 

Region 13 contains very faint outermost filaments in which we detect only 
$\sim$700 counts. Our best-fit NEI plane shock model fit for this regional 
spectrum shows generally similar results to those for other ISM regions.
However, because of significantly poorer photon count statistics in this
regional spectrum, uncertainties on the best-fit values of the model 
parameters are large, and the measurements are less reliable for this 
region. Thus, we do not present the quantitative measurements of our 
fitted spectral model parameters for this region. Regions 5 and 10 also 
include relatively lower count statistics of $\sim$2000 counts. Our 
spectral model fits for these regions do not allow a unique solution, and 
result in large uncertainties on the best-fit values. Spectra from these 
regions can be equally described by a few different NEI plane shock model 
fits after fixing some parameters (e.g., $N_{H, LMC}$ and $kT$) at the 
mean values of the neighboring regions. Since results from those model 
fits are consistent (within uncertainties), we present average values of 
the best-fit parameters between these alternative model fits for these two 
regions. These results are in agreement with those from other shocked ISM 
regions. We summarize these results in Tables~\ref{tbl:tab2} \& \ref{tbl:tab3}.

\section{\label{sec:disc} DISCUSSION}

\begin{figure}[]
\figurenum{1}
\centerline{\includegraphics[angle=0,width=\textwidth]{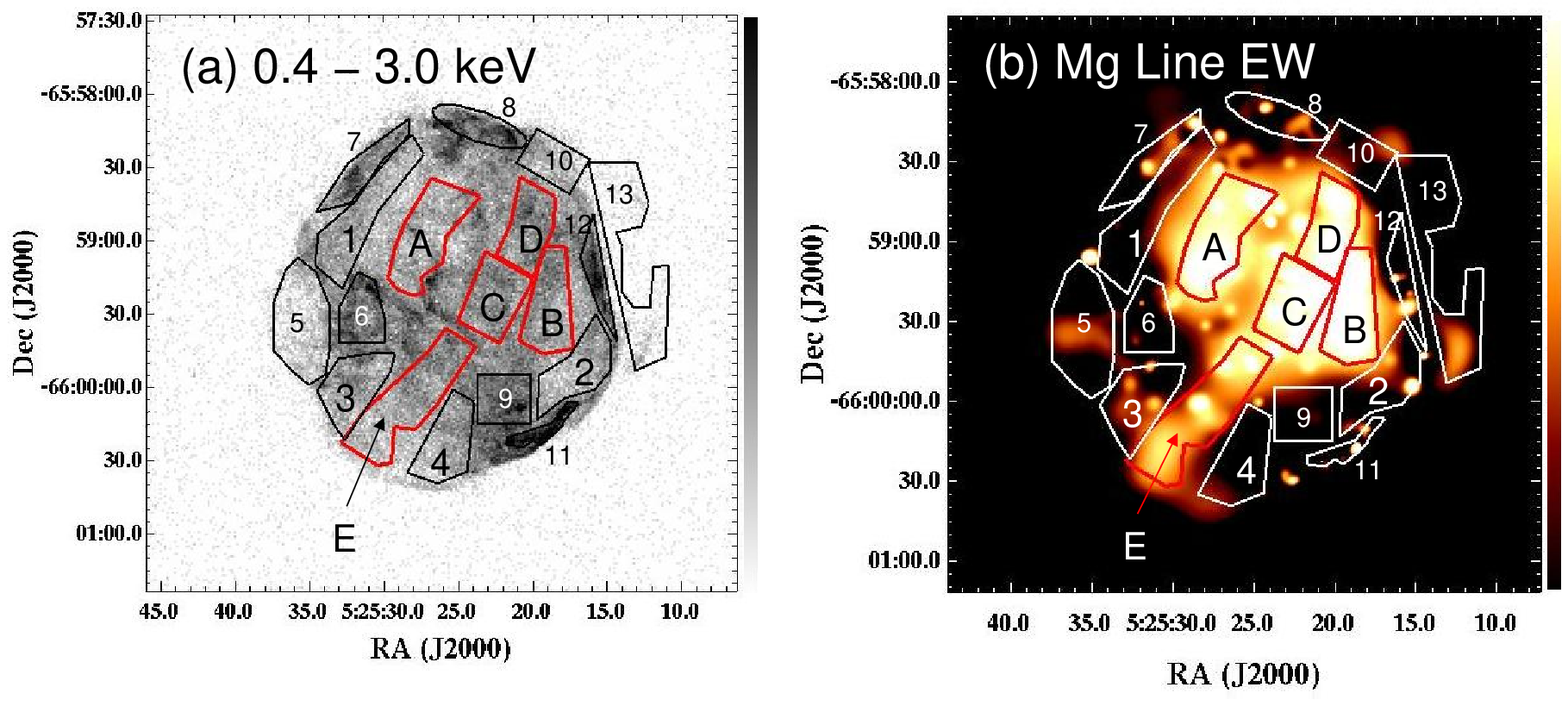}}
\figcaption[]{(a) A broadband grey-scale {\it Chandra} ACIS image of N49B. The 
image has been binned by $\sim$1$^{\prime\prime}$ $\times$ 1$^{\prime\prime}$
pixels. (b) A false-color Mg line EW image of N49B. To create this map we used 
the ACIS subband images (see the text) with the original pixel size (0$\farcs$492), 
and adaptively smoothed them for the purposes of display. In (a) and (b), regions 
of our spectral analysis are overlaid.
\label{fig:fig1}}
\end{figure}

\begin{figure}[]
\figurenum{2}
\centerline{\includegraphics[angle=0,width=\textwidth]{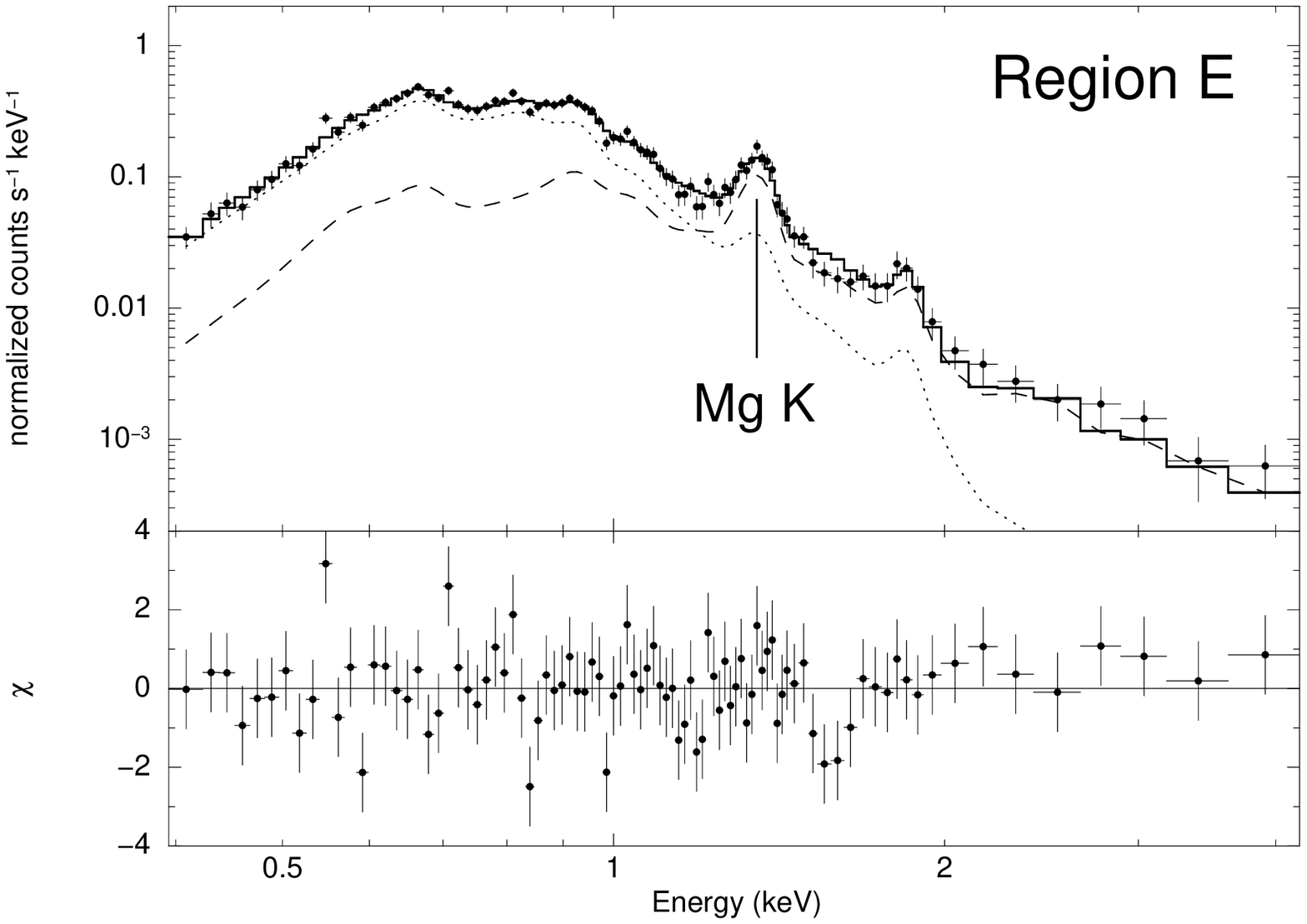}}
\figcaption[]{The ACIS spectrum extracted from region E (as marked in 
Figure~\ref{fig:fig1}). The best-fit two-component plane-shock model (solid 
curve) is overlaid. The soft (dotted) and hard (dashed) components of the 
best-fit plane shock model are overlaid. The bottom plot shows residuals from 
the best-fit model. 
\label{fig:fig2}}
\end{figure}

\begin{figure}[]
\figurenum{3}
\centerline{\includegraphics[angle=0,width=\textwidth]{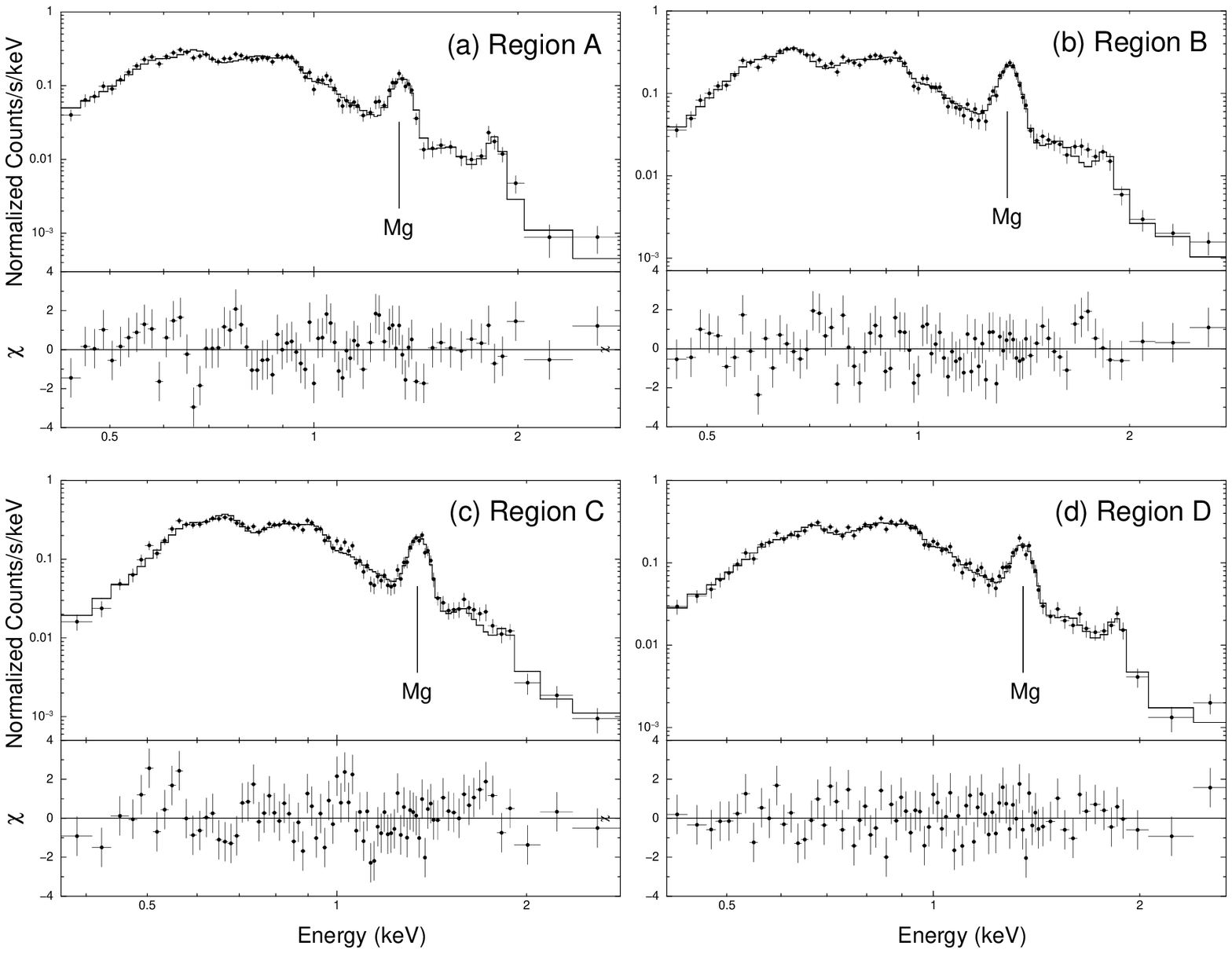}}
\figcaption[]{The ACIS spectra extracted from Mg-overabundant regions (regions A, B, 
C, and D, as marked in Figure~\ref{fig:fig1}). The best-fit plane-shock model is 
overlaid in each panel. In each panel the bottom plot shows residuals from the 
best-fit model. 
\label{fig:fig3}}
\end{figure}

\begin{figure}[]
\figurenum{4}
\centerline{\includegraphics[angle=0,width=\textwidth]{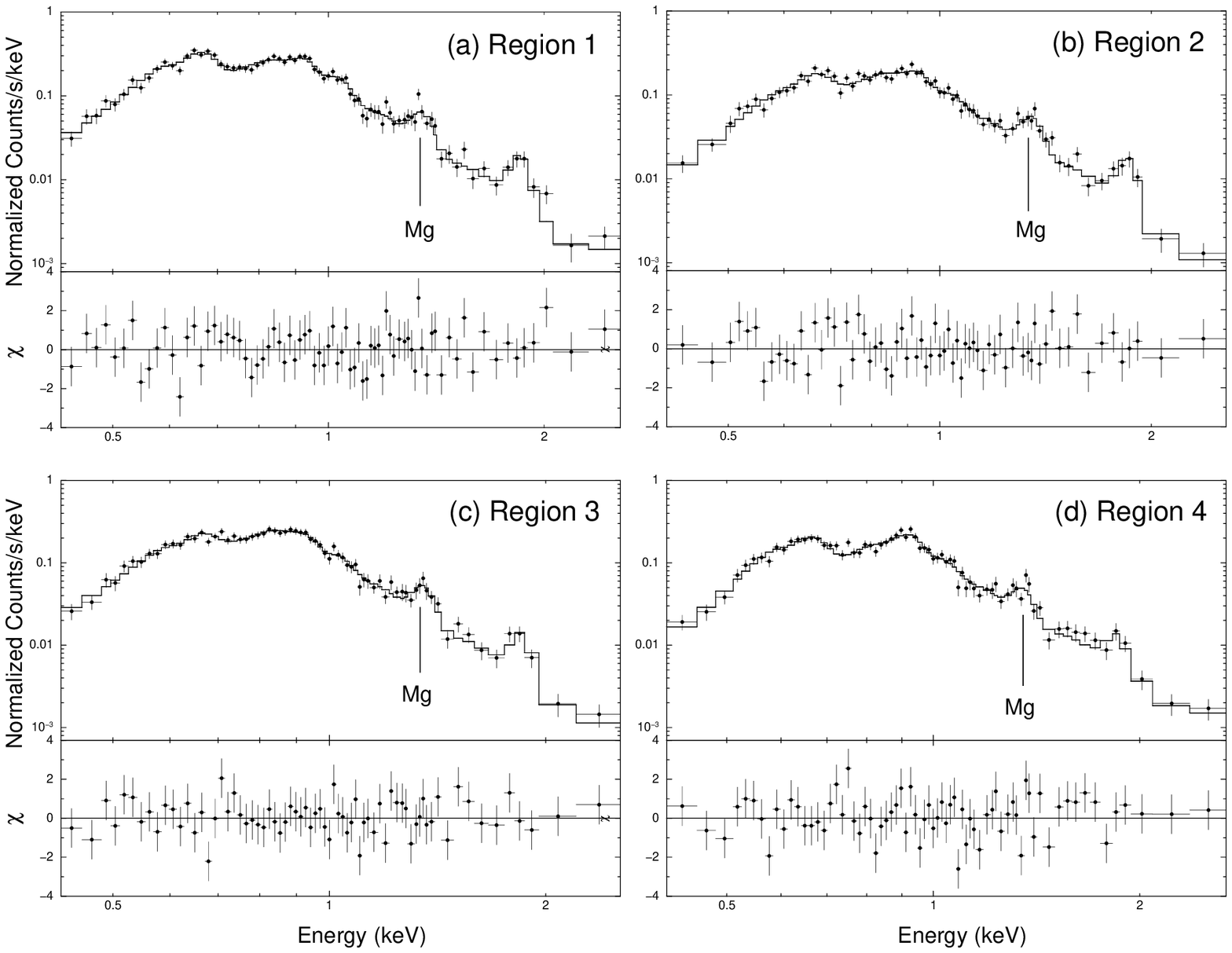}}
\figcaption[]{Four example ACIS spectra extracted from shocked ISM regions 
(regions 1, 2, 3, and 4, as marked in Figure~\ref{fig:fig1}). The best-fit 
plane-shock model is overlaid in each panel. In each panel the bottom plot 
shows residuals from the best-fit model. 
\label{fig:fig4}}
\end{figure}

\begin{figure}[]
\figurenum{5}
\centerline{\includegraphics[angle=0,width=\textwidth]{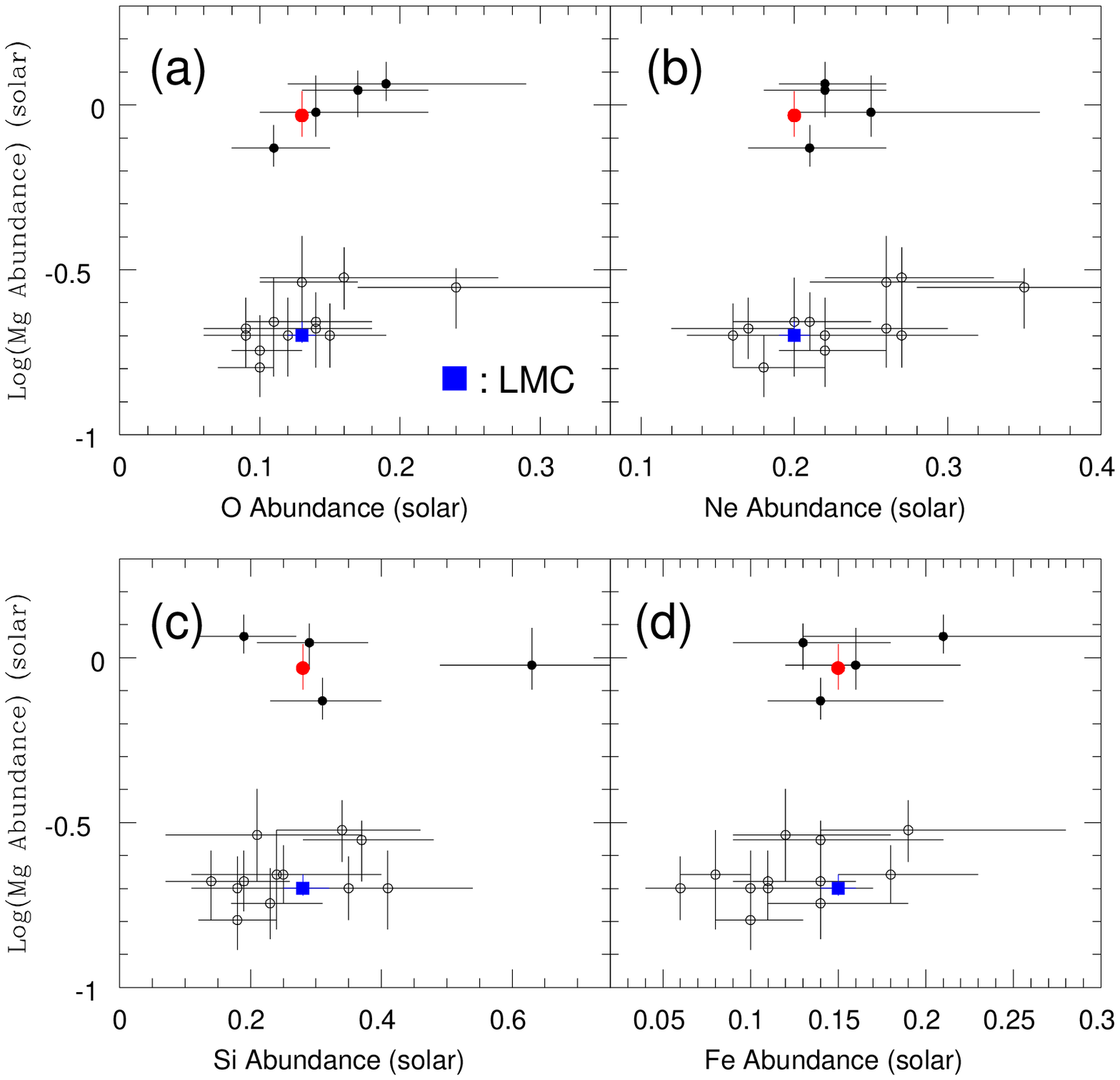}}
\figcaption[]{Comparisons of Mg abundance with those of O, Ne, Si, and Fe
for sub-regions of N49B. Black filled circles are regions A -- D. The red 
filled circle shows the hard component of region E. Open circles are
regions 1 -- 12. The blue square is the LMC abundance \citep{sche16}.
\label{fig:fig5}}
\end{figure}

\begin{figure}[]
\figurenum{6}
\centerline{\includegraphics[angle=0,width=\textwidth]{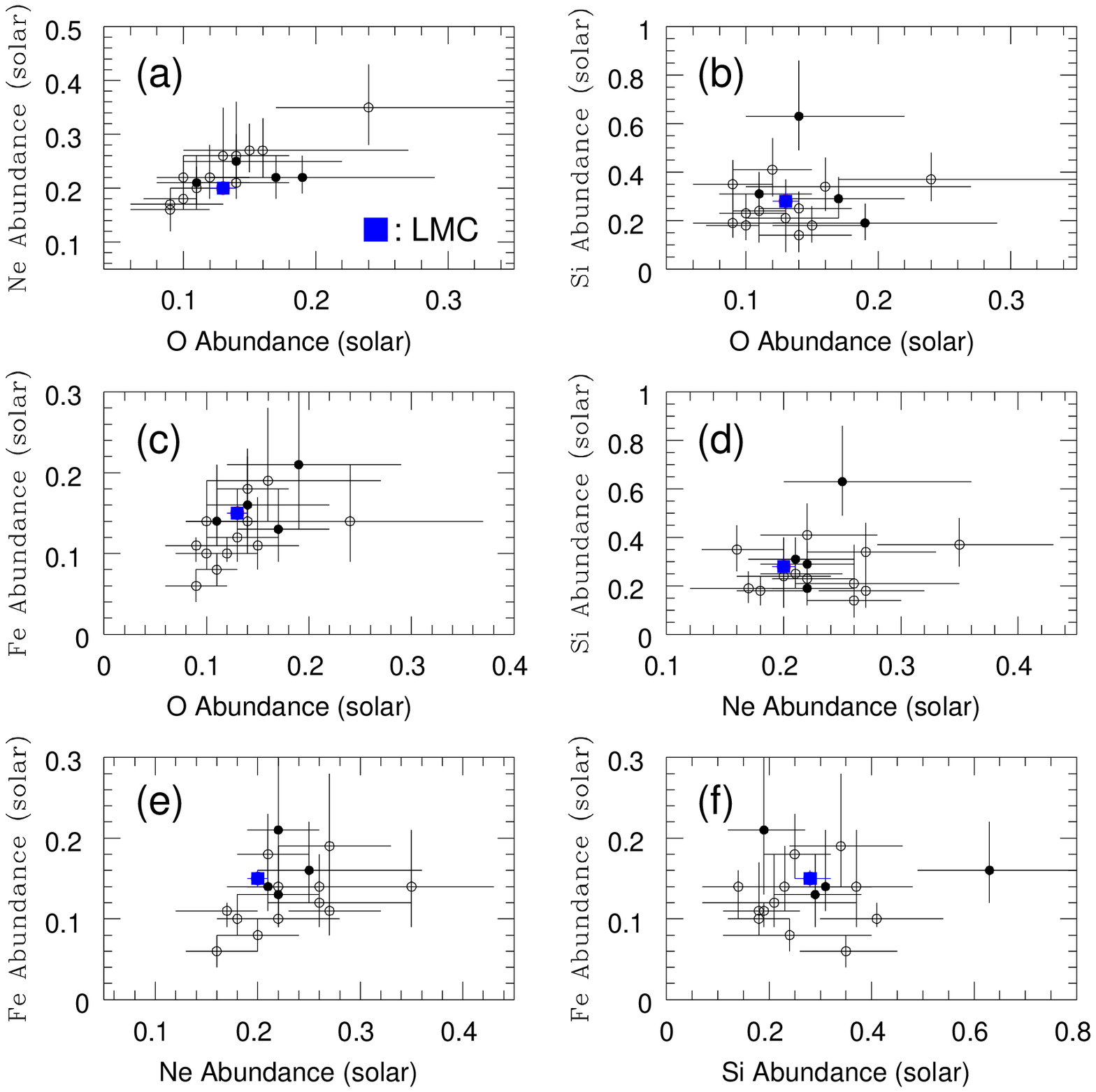}}
\figcaption[]{Comparisons of measured elemental abundances (except for Mg)
for sub-regions of N49B. Black filled circles are regions A -- D, and open 
circles are regions 1 -- 12. The blue square is the LMC abundance \citep{sche16}.
The LMC abundances are assumed for region E (see the text).
\label{fig:fig6}}
\end{figure}

\begin{figure}[]
\figurenum{7}
\centerline{\includegraphics[angle=0,width=\textwidth]{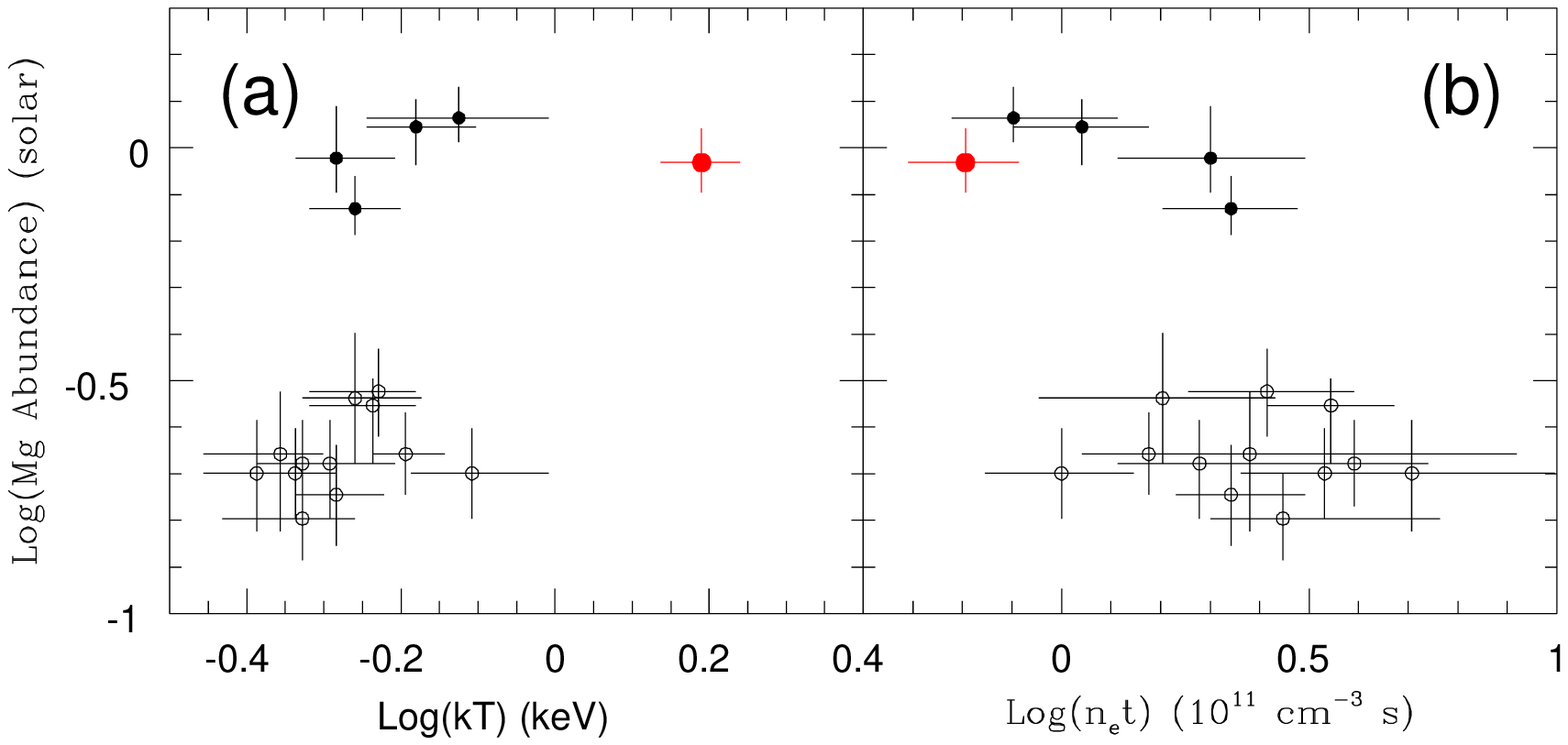}}
\figcaption[]{Comparisons of Mg abundance with electron temperature and
ionization timescale for sub-regions of N49B. Black filled circles are 
regions A -- D. The red filled circle is the hard component of region E. 
Open circles are regions 1 -- 12.
\label{fig:fig7}}
\end{figure}

Our spectral analysis of the observed X-ray spectra for a number of
subregions in N49B shows that Mg is clearly overabundant in regions A -- E
when compared to that in ISM-dominated regions (Figure~\ref{fig:fig5}). In 
contrast to Mg the estimated abundances for other metal elements (O, Ne, 
Si, and Fe) are consistent with the LMC values throughout the entire SNR 
(Figures~\ref{fig:fig5} \& \ref{fig:fig6}). This Mg-only enhancement is 
consistently observed for region E in addition to regions A -- D. The 
electron temperature and ionization timescale show similar ranges between 
the ejecta and ISM regions (probably except for region E, 
Figure~\ref{fig:fig7}), suggesting that the observed enhancements in 
the Mg line EW in N49B are primarily due to the Mg overabundance. It is 
remarkable that X-ray emission from the shocked Mg-rich ejecta gas 
apparently extends from the central ejecta regions (regions A -- D) 
out to the southeastern outermost boundary of the SNR. This extended 
Mg-rich ejecta gas is confined in an elongated region (region E in 
Figure~\ref{fig:fig1}), presumably with a cylindrical geometry. The Mg 
abundance in this region is estimated to be higher than the LMC abundance 
by a factor of $\sim$5, which is consistent with the Mg overabundance 
detected in the central regions of this SNR. It is interesting to find 
that the Mg-rich ejecta gas in region E shows a significantly higher 
temperature ($kT$ $\sim$ 1.6 keV) than that of the central ejecta gas 
($kT$ $\sim$ 0.6 keV) (Figure~\ref{fig:fig7}). The Mg overabundance and 
high electron temperature in region E are persistent between the northern 
and southern halves of region E (although larger uncertainties are implied 
for the best-fit values due to lower count statistics in these half-regions). 
Thus, region E likely represents a {\it continuous outflow} of the hot 
Mg-rich ejecta gas channeled from the SNR center to the outermost boundary 
of the SNR.   

P03 estimated a large mass of the Mg-rich ejecta ($\ga$1 $M_{\odot}$) 
for N49B, extrapolating the measured Mg abundance in a small Mg-enhanced 
region near the SNR center (roughly corresponding to the southern half 
of region A) to the {\it entire} Mg-enhanced region in the SNR center 
(a spherical volume with a radius of $\sim$0$\farcm$7). Although it was 
a crude approximation, this large Mg mass suggested a very massive 
progenitor star ($M$ $>$ 25 $M_{\odot}$) for N49B (P03). Based on our 
extensive spectral analysis of the entire Mg-rich regions, we revisit our 
Mg ejecta mass estimate to better constrain it. We estimate the emission 
volume for each of regions A -- E, and consider their sum as the {\it total} 
volume of the Mg-enriched regions. At the distance of $d$ = 50 kpc the 
projected angular areas correspond to $\sim$4.8 pc $\times$ 11.6 pc, 
$\sim$4.8 pc $\times$ 9.7 pc, $\sim$6.3 pc $\times$ 6.3 pc, and $\sim$4.4 
pc $\times$ 7.8 pc for regions A, B, C, and D, respectively. Assuming 
$\sim$5 pc (a physical scale similar to that for the {\it mean} of the 
projected angular sizes of regions A, B, and D) for the path-length along 
the line of sight for these regions, we estimate emission volumes of $V_A$ 
$\sim$ 11.5 $\times$ 10$^{57}$ $f$ cm$^3$, $V_B$ $\sim$ 9.7 $\times$ 
10$^{57}$ $f$ cm$^3$, and $V_D$ $\sim$ 7.1 $\times$ 10$^{57}$ $f$ cm$^3$, 
where $f$ is the X-ray emitting volume filling factor. Since region C is 
closer to the center of the SNR (whose overall geometry is presumed to be 
roughly spherical), we assumed a relatively longer path-length of $\sim$10 
pc for region C (a physical scale similar to the radius of the SNR). Then, 
we estimate the emission volume for region C to be $V_C$ $\sim$ 11.7 $\times$ 
10$^{57}$ $f$ cm$^3$. Using the best-fit EM and Mg abundances 
(Tables~\ref{tbl:tab1} \& \ref{tbl:tab2}), we estimate the electron density 
$n_e$ $\sim$ 0.026, 0.031, 0.023, and 0.035 $f^{-{1\over2}}$ cm$^{-3}$ for 
regions A, B, C, and D, respectively. Although our estimates for regional 
emission volumes include some uncertainties (probably up to by a factor of 
a few, depending on the assumed geometry and physical scales of individual
regions), it would not significantly affect our density estimates because 
$n_e$ $\propto$ $V^{1\over2}$. In these calculations we assumed all electrons 
are liberated from He-like Mg ion ($n_e$ $\sim$ 10$n_{\rm Mg}$, where 
$n_{\rm Mg}$ is the Mg ion density) for a simple ``pure-ejecta'' case.
Assuming the dominant isotope $^{24}$Mg, the total Mg ejecta mass for regions
A -- D is $M_{\rm Mg}$ = 24 $m_p$ $n_{\rm Mg}$ $V$ (where $m_p$ is the proton 
mass and the total volume $V$ = $V_A$ + $V_B$ + $V_C$ + $V_D$ $\sim$ 4 
$\times$ 10$^{58}$ $f$ cm$^3$) $\sim$ 2.4 $f^{1\over2}$ $M_{\odot}$. If we 
assumed a simple spherical volume ($V$ $\sim$ 1.3 $\times$ 10$^{59}$ $f$ 
cm$^3$ for an angular radius of 0$\farcm$7, corresponding to $\sim$10 pc 
at $d$ = 50 kpc, as adopted by P03) for the central Mg-rich ejecta region 
and the mean Mg ion density of $n_{\rm Mg}$ $\sim$ 0.003 $f^{-{1\over2}}$ 
cm$^{-3}$ for regions A -- D, a larger mass of $M_{\rm Mg}$ $\sim$ 8 
$f^{1\over2}$ $M_{\odot}$ is estimated. A conservative lower limit of the
ejecta mass $M$ $\sim$ 0.1 $f^{1\over2}$ $M_{\odot}$ may be estimated by 
replacing the Mg ion mass with $m_p$ for an emission volume of $V$ $\sim$ 
4 $\times$ 10$^{58}$ $f$ cm$^3$.

We estimate the Mg ejecta mass in region E in a similar method. We assume 
a cylindrical volume (with $\sim$2.5 pc in radius and $\sim$16 pc in length
for $d$ = 50 kpc) for this region based on the apparently elongated morphology 
in projection. The estimated emission volume is $V_E$ $\sim$ 8.5 $\times$ 
10$^{57}$ $f$ cm$^3$. Based on the best-fit EM and Mg abundance for the hard 
component plane shock model (Table~\ref{tbl:tab1}), we estimate $n_e$ $\sim$ 
0.015 $f^{-{1\over2}}$ cm$^{-3}$ and the Mg mass $M_{\rm Mg}$ $\sim$ 0.26 
$f^{1\over2}$ $M_{\odot}$ (for the pure-ejecta assumption). This is $\sim$10\% 
of the total estimated $M_{\rm Mg}$ for the entire Mg-rich ejecta region 
(regions A -- E). For an alternative spectral modeling of the region E 
spectrum, in which we fitted the O, Ne, Si, and Fe abundances in addition 
to the Mg abundance (see Section~\ref{subsec:jet}), our estimated Mg ejecta 
mass is nearly the same. For the case of the Mg-rich ejecta with a low 
temperature in region E (based on a plane shock + PL model fit, 
Section~\ref{subsec:jet}) a larger $n_e$ is estimated, and the implied Mg 
mass would also be larger (by a factor of $\sim$3).   
Our estimated value for the Mg-rich ejecta mass for the central ejecta region 
is generally lower than that estimated by P03. We note that our estimate is 
based on a more extensive regional spectral analysis of the {\it entire} 
Mg-ejecta region, and thus we conclude our estimate to be more realistic 
than the previous value. In fact, even the lower limit for the estimated Mg 
ejecta mass by P03 was significantly higher than the largest Mg mass available 
in the core-collapse SN nucleosynthesis model calculations \citep{thie96}, 
whereas our new lower limit is generally in better agreement with model 
calculations \citep{thie96,nomo06}. Nonetheless, to produce such a large 
amount of Mg from a core-collapse explosion of a progenitor star with an 
LMC-like metallicity, the progenitor masses of $M$ $<$ 20 $M_{\odot}$ may be 
ruled out based on standard SN nucleosynthesis calculations \citep{nomo06}. 

\begin{figure}[]
\figurenum{8}
\centerline{\includegraphics[angle=0,width=\textwidth]{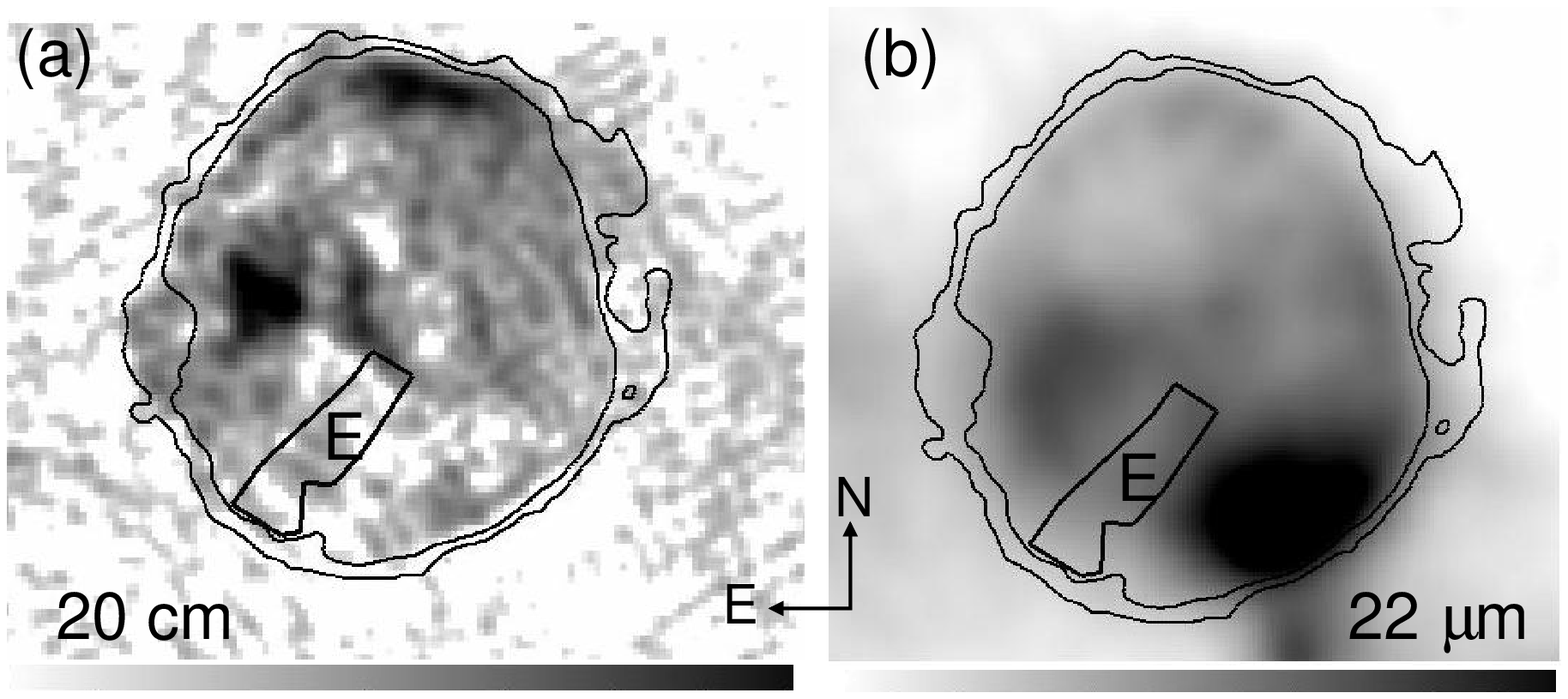}}
\figcaption[]{(a) A grey-scale ATCA 20 cm image of N49B. The angular 
resolution (half-power beamwidth) is circular 6$^{\prime\prime}$. (b) 
A grey-scale 22 $\mu$m image of N49B from the public archive of the 
{\it WISE} all-sky survey data. The angular resolution is 
12$^{\prime\prime}$. In (a) and (b) contours of the 0.3 -- 3.0 keV ACIS 
image of N49B are overlaid to show the outer boundary of the SNR in the 
X-ray band. Region E is marked in (a) and (b).
\label{fig:fig8}}
\end{figure}

The southeastern elongation of the Mg-rich ejecta might have been caused 
by a significantly non-uniform ambient medium, which might have allowed 
the ejecta gas to expand into the particular channel in the southeastern 
direction. The consistence in the ejecta composition (i.e., Mg-rich) between 
region E and the central ejecta regions (regions A -- D) may support this 
interpretation. The radio and mid-infrared (MIR) images of N49B show the 
{\it lack} of interstellar gas and dust in the south-southeastern regions 
of N49B \citep{dick98,will06}, generally corresponding to region E 
(Figure~\ref{fig:fig8}). This coincidence is more pronounced in the MIR 
image than in the radio map. The northeastern and western regions of the 
SNR (roughly corresponding to regions A and B, respectively) also show a 
similarly low MIR emission with a correspondingly high Mg abundance. This 
general anti-correlation between the MIR intensity and the Mg abundance is 
consistent with this scenario. In this interpretation the ejecta gas in 
region E might have been shocked multiple times to a higher temperature 
than that of the bulk of the ejecta gas, because it is confined within a 
relatively small, low-density volume surrounded by a high density medium. 
X-ray emission from multi-phase hot gas caused by reheating of the 
low-density gas in inter-cloud regions has been proposed for some other 
SNRs interacting with clumpy clouds \citep[e.g.,][]{hest94,park03a}. 
However, it is unclear whether the particular geometry and the ambient 
density structure of region E and its surrounding regions would
suffice such a condition. A cylindrical geometry for region E would be 
effective to secure a small volume of the low-density gas, and thus for 
the reheating. The origin for the formation of such a cylindrical {\it 
tunnel} with a large scale ($\sim$5 pc $\times$ 16 pc) in the ambient 
medium is unclear. Although it is less pronounced, the northwestern
outer boundary of N49B (roughly corresponding to region 10) shows a
low MIR intensity similar to that in region E\footnote{This is more 
evident in the 24$\mu$m {\it Spitzer} image as shown in Figure~1 
of Williams et al. (2006).}. A similar extent of the Mg-rich ejecta
might be expected in this region, which is not the case. While the 
density-dependence of the radio flux is more complex than the MIR dust 
emission, the anti-correlation of the radio intensity against the Mg 
abundance is less clear. Thus, while it is a plausible scenario, this 
interpretation remains speculative. Detailed hydrodynamic simulations 
would be necessary to test this non-uniform medium interpretation for 
the observed elongated morphology of the hot Mg-rich ejecta in N49B. 

Considering that a relatively high explosion energy ($E_0$ $\sim$ 4 $\times$
10$^{51}$ erg) was estimated for N49B \citep{hugh98} (albeit this $E_0$ might
have been somewhat overestimated as discussed by those authors), we entertain an alternative
scenario for the origin of the elongated Mg-rich ejecta in this SNR.
Core-collapse SN explosions are considered to be more asymmetric than Type Ia
thermonuclear SN, and such asymmetric explosions may be imprinted in their SNRs
\citep[e.g.,][]{lope14,gonz14}. Models for SN explosions driven by an energetic
jet (powered by accretion onto the central compact remnant, Maeda \& Nomoto
2003) is particularly intriguing. This {\it jet-driven} SNR scenario has been
argued for to explain the observed elongated X-ray morphology of the metal-rich
ejecta in the Galactic SNR W49B \citep{keoh07,lope13,gonz14} and the SNR 
0104--72.3 in the Small Magellanic Cloud \citep{lope14}. The elongated morphology 
for the southeastern metal-rich ejecta in N49B is generally consistent with the 
fossil of the jet-like outflow similar to those seen in W49B and 0104--72.3. The 
estimated ejecta mass fraction ($\sim$10\% of the total ejecta) is also consistent
with that predicted by the jet-driven SN models \citep{maed03}.  However, the 
elongated ejecta emission in N49B is enriched in Mg in contrast to the Fe-rich 
ejecta found in W49B \citep{lope13} and 0104--72.3 \citep{lee11,lope14}. In fact, 
the jet-driven SN models predict Fe- and Ni-rich (created in the deepest core of 
the SN) outflows, while lighter metals such as O, Ne, and Mg would show lower 
expansion velocities, and thus would likely be located near the central regions 
of the SNR \citep{maed03}. Formation of an energetic bi-polar outflow from the 
accretion onto the central compact remnant (probably a black hole) is expected 
in the jet-driven explosion \citep{maed03}. If region E in N49B were the relic 
of a jet-like ejecta outflow toward the southeast, one may expect a similar 
structure on the opposite side of the SNR. We find no clear evidence for such 
a counter ejecta outflow toward the northwestern outer boundary of N49B 
(Figure~\ref{fig:fig1}b): i.e., the likely location for such a feature would 
be region 10 in which metal overabundance is not detected. Thus, alhough
it may not be ruled out, the interpretation as a jet-driven SN for the origin 
of N49B is in question. Nonetheless, we note that the central compact remnant 
has not been detected in N49B (P03). The estimated upper limit (3$\sigma$) 
on the X-ray luminosity (in the 0.5 -- 5 keV band) for the putative compact 
remnant is $L_X$ $\sim$ 2 $\times$ 10$^{33}$ erg s$^{-1}$ (P03). Although this 
luminosity upper limit is not particularly constraining (which still allows 
a non-detection of a normal neutron star due to a long distance to the LMC), 
it is not inconsistent with the creation of a black hole from a jet-driven 
SN explosion of a very massive star.

With the currently available data the origin of the elongated Mg-rich ejecta in 
SNR N49B remains elusive. Further observational and theoretical studies are needed 
to help answer the fundamental questions related to this peculiar core-collapse SNR: 
Why is the metal-rich ejecta in this SNR enriched only in Mg, and what caused the 
southeastern elongation of the observed Mg-rich ejecta? Deep high-resolution X-ray 
observations would be necessary to study the detailed spatial and spectral structures 
of the Mg-rich ejecta in N49B and thus its origin, particularly to reveal the true 
nature of the hard component emission in region E. Deep high-resolution radio and 
IR observations would be needed to study the detailed ambient structure to help 
unveil the origin of the southeastern extent of the Mg-rich ejecta. Such deep 
observations would also help to place observational constraints on the nature of 
the central compact remnant created in N49B, which is essential to understanding 
of the observational properties of this peculiar core-collapse SNR.

\acknowledgments

The authors thank the anonymous referee for her/his constructive criticisms
that helped improving this manuscript. This work has been supported in part 
by NASA through the {\it Chandra} grant AR0-11008A issued by the {\it Chandra} 
X-ray Observatory Center, which is operated by the Smithsonian Astrophysical 
Observatory. We thank John Dickel for providing the 20 cm image of N49B, taken 
by the Australian Telescope Compact Array (ATCA). This publication makes use of 
data products from the {\it Wide-field Infrared Survey Explorer} ({\it WISE}), 
which is a joint project of the University of California, Los Angeles, and the 
Jet Propulsion Laboratory/California Institute of Technology, funded by NASA.

\clearpage

\begin{deluxetable}{cccccc}
\footnotesize
\tablecaption{Results from two-component NEI shock model fit for Region E Spectrum
\label{tbl:tab1}}
\tablewidth{0pt}
\tablehead{\colhead{} & \colhead{$N_{H,LMC}$\tablenotemark{a}} & \colhead{$kT$} & 
\colhead{$\tau$} & \colhead{Mg} & \colhead{EM} \\
\colhead{} & \colhead{(10$^{21}$ cm$^{-2}$)} & \colhead{(keV)} & \colhead{(10$^{11}$ cm$^{-3}$ s)} 
&  \colhead{(solar)} & \colhead{(10$^{58}$ cm$^{-3}$)} } 
\startdata
Soft & 2.7$^{+0.3}_{-0.2}$ & 0.27$\pm$0.01 & $>$50.0 & 0.20 (fixed\tablenotemark{b}) 
& 12.2$^{+1.6}_{-1.0}$ \\
Hard & 2.7$^{+0.3}_{-0.2}$ & 1.55$^{+0.19}_{-0.18}$ & 0.6$^{+0.2}_{-0.2}$ 
& 0.93$^{+0.17}_{-0.13}$ & 0.5$\pm$0.1 \\
\enddata
\tablecomments{1$\sigma$ errors are shown. $\chi^2$/$\nu$ = 85.1/84.}
\tablenotetext{a}{A common $N_{H,LMC}$ for both of the soft and hard components is assumed.}
\tablenotetext{b}{The Mg abundance for the soft component as well as O, Ne, Si, S, and Fe
abundances for both components are fixed at the LMC values \citep{sche16}.}
\end{deluxetable}

\begin{deluxetable}{cccccc}
\footnotesize
\tablecaption{Spectral Parameters from NEI shock model fits for Regional Spectra of N49B
\label{tbl:tab2}}
\tablewidth{0pt}
\tablehead{\colhead{Regions} & \colhead{$N_{H,LMC}$} & \colhead{$kT$} & 
\colhead{$\tau$} & \colhead{EM} &  \colhead{$\chi^2$/$\nu$} \\
\colhead{} & \colhead{(10$^{21}$ cm$^{-2}$)} & \colhead{(keV)} & \colhead{(10$^{11}$ cm$^{-3}$ s)} 
& \colhead{(10$^{58}$ cm$^{-3}$)} & \colhead{} } 
\startdata
A & 1.4$^{+0.6}_{-0.8}$ & 0.52$^{+0.10}_{-0.06}$ & 2.0$^{+1.1}_{-0.7}$ & 2.0$^{+1.1}_{-0.9}$ &
85.0/68 \\
B & 2.7$^{+0.8}_{-0.6}$ & 0.66$^{+0.13}_{-0.09}$ & 1.1$^{+0.4}_{-0.3}$ & 2.1$^{+1.2}_{-0.8}$ &
81.8/75 \\ 
C & 2.0$^{+0.9}_{-0.6}$ & 0.75$^{+0.23}_{-0.18}$ & 0.8$^{+0.5}_{-0.2}$ & 1.4$^{+1.7}_{-0.7}$ &
100.6/76 \\
D & 2.5$^{+0.7}_{-0.4}$ & 0.55$^{+0.08}_{-0.07}$ & 2.2$^{+0.8}_{-0.6}$ & 3.1$^{+1.8}_{-1.2}$ &
70.0/74 \\
1 & 1.7$^{+0.5}_{-0.4}$ & 0.58$^{+0.08}_{-0.10}$ & 3.5$^{+1.2}_{-0.9}$ & 2.2$^{+1.4}_{-0.6}$ &
74.2/70 \\
2 & 3.9$^{+1.0}_{-0.6}$ & 0.46$^{+0.06}_{-0.10}$ & 3.4$^{+6.5}_{-1.1}$ & 4.6$^{+5.6}_{-1.4}$ &
59.5/62 \\
3 & 1.8$^{+0.8}_{-0.5}$ & 0.59$^{+0.07}_{-0.11}$ & 2.6$^{+1.3}_{-0.8}$ & 1.8$^{+1.5}_{-0.6}$ &
48.8/62 \\
4 & 2.9$\pm$0.5 & 0.78$^{+0.20}_{-0.13}$ & 1.0$^{+0.4}_{-0.3}$ & 1.4$^{+0.7}_{-0.5}$ &
71.9/63 \\
5 & 1.8$^{+0.5}_{-0.6}$ & 0.55$^{+0.12}_{-0.08}$ & 1.6$^{+1.1}_{-0.7}$ & 1.1$\pm$0.4 &
59.2/49 \\
6 & 1.2$^{+0.6}_{-0.4}$ & 0.47$^{+0.05}_{-0.06}$ & 3.9$^{+1.6}_{-1.2}$ & 4.7$^{+2.6}_{-1.4}$ &
93.2/73 \\
7 & 2.3$^{+0.5}_{-0.6}$ & 0.41$^{+0.10}_{-0.06}$ & 5.1$^{+4.8}_{-2.1}$ & 4.4$^{+3.4}_{-2.2}$ &
73.8/59 \\
8 & 1.1$^{+0.6}_{-0.3}$ & 0.51$^{+0.11}_{-0.07}$ & 1.9$^{+0.8}_{-0.6}$ & 2.8$^{+1.6}_{-1.2}$ &
46.9/65 \\
9 & 1.5$^{+0.4}_{-0.5}$ & 0.64$^{+0.08}_{-0.06}$ & 1.5$^{+0.6}_{-0.4}$ & 2.7$^{+0.8}_{-0.7}$ &
79.3/74 \\ 
10 & 1.6 (fixed\tablenotemark{a}) & 0.44$^{+0.06}_{-0.09}$ & 2.4$^{+5.9}_{-0.9}$ & 1.5$^{+1.0}_{-0.4}$ &
44.4/44 \\ 
11 & 3.5$^{+1.0}_{-0.6}$ & 0.47$^{+0.08}_{-0.10}$ & 2.8$^{+3.0}_{-0.8}$ & 5.7$^{+7.2}_{-2.1}$ &
95.5/70 \\
12 & 2.1$^{+0.5}_{-0.4}$ & 0.52$^{+0.0.08}_{-0.06}$ & 2.2$^{+0.9}_{-0.5}$ & 3.4$^{+1.7}_{-1.2}$ &
71.3/67 \\
\enddata
\tablecomments{1$\sigma$ errors are shown.}
\tablenotetext{a}{$N_{H,LMC}$ for region 10 is not constrained and we fix it at
the mean value of nearby regions 8 and 12.}
\end{deluxetable}

\begin{deluxetable}{cccccc}
\footnotesize
\tablecaption{Best-Fit Metal Abundances for Regional Spectra of N49B
\label{tbl:tab3}}
\tablewidth{0pt}
\tablehead{\colhead{Regions} & \colhead{O\tablenotemark{a}} & \colhead{Ne\tablenotemark{a}} 
& \colhead{Mg\tablenotemark{a}} & \colhead{Si\tablenotemark{a}} & \colhead{Fe\tablenotemark{a}} \\
\colhead{} &  \colhead{} & \colhead{} & \colhead{} & \colhead{} & \colhead{} } 
\startdata
A & 0.14$^{+0.0.08}_{-0.04}$ & 0.25$^{+0.11}_{-0.05}$ & 0.95$^{+0.28}_{-0.15}$ & 0.63$^{+0.23}_{-0.14}$ &
0.16$^{+0.06}_{-0.04}$ \\
B & 0.17$^{+0.05}_{-0.04}$ & 0.22$\pm$0.04 & 1.11$^{+0.16}_{-0.19}$ & 0.29$^{+0.09}_{-0.08}$ &
0.13$^{+0.05}_{-0.04}$ \\ 
C & 0.19$^{+0.10}_{-0.07}$ & 0.22$^{+0.04}_{-0.03}$ & 1.16$^{+0.19}_{-0.13}$ & 0.19$^{+0.08}_{-0.07}$ &
0.21$^{+0.18}_{-0.08}$ \\
D & 0.11$^{+0.04}_{-0.03}$ & 0.21$^{+0.05}_{-0.04}$ & 0.74$^{+0.13}_{-0.09}$ & 0.31$^{+0.09}_{-0.08}$ &
0.14$^{+0.07}_{-0.03}$ \\
1 & 0.24$^{+0.13}_{-0.07}$ & 0.35$^{+0.08}_{-0.07}$ & 0.28$^{+0.04}_{-0.07}$ & 0.37$^{+0.11}_{-0.09}$ &
0.14$^{+0.07}_{-0.05}$ \\
2 & 0.09$\pm$0.03 & 0.16$^{+0.04}_{-0.03}$ & 0.20$^{+0.05}_{-0.04}$ & 0.35$^{+0.10}_{-0.09}$ &
0.06$\pm$0.02 \\
3 & 0.16$^{+0.11}_{-0.06}$ & 0.27$^{+0.06}_{-0.05}$ & 0.30$^{+0.07}_{-0.06}$ & 0.34$^{+0.12}_{-0.10}$ &
0.19$^{+0.09}_{-0.05}$ \\
4 & 0.15$^{+0.04}_{-0.03}$ & 0.27$^{+0.05}_{-0.04}$ & 0.20$^{+0.05}_{-0.04}$ & 0.18$^{+0.08}_{-0.07}$ &
0.11$^{+0.06}_{-0.03}$ \\
5 & 0.13$^{+0.04}_{-0.03}$ & 0.26$^{+0.09}_{-0.05}$ & 0.29$^{+0.11}_{-0.08}$ & 0.21$^{+0.16}_{-0.14}$ &
0.12$^{+0.06}_{-0.03}$ \\
6 & 0.09$^{+0.04}_{-0.03}$ & 0.17$^{+0.03}_{-0.05}$ & 0.21$^{+0.05}_{-0.04}$ & 0.19$^{+0.07}_{-0.06}$ &
0.11$^{+0.01}_{-0.02}$ \\
7 & 0.12$\pm$0.03 & 0.22$^{+0.06}_{-0.04}$ & 0.20$^{+0.06}_{-0.05}$ & 0.41$^{+0.13}_{-0.11}$ &
0.10$^{+0.02}_{-0.01}$ \\
8 & 0.14$^{+0.04}_{-0.03}$ & 0.26$\pm$0.04 & 0.21$\pm$0.05 & 0.14$^{+0.08}_{-0.07}$ & 
0.14$^{+0.02}_{-0.04}$ \\
9 & 0.14$^{+0.04}_{-0.03}$ & 0.21$^{+0.04}_{-0.03}$ & 0.22$^{+0.05}_{-0.04}$ & 0.25$^{+0.07}_{-0.06}$ &
0.18$^{+0.05}_{-0.03}$ \\
10 & 0.11$\pm$0.02 & 0.20$\pm$0.04 & 0.22$^{+0.08}_{-0.07}$ & 0.24$^{+0.16}_{-0.13}$ & 0.08$\pm$0.02 \\
11 & 0.10$^{+0.01}_{-0.03}$ & 0.18$^{+0.04}_{-0.02}$ & 0.16$^{+0.04}_{-0.03}$ & 0.18$\pm$0.06 &
0.10$^{+0.03}_{-0.02}$ \\
12 & 0.10$^{+0.03}_{-0.02}$ & 0.22$^{+0.04}_{-0.03}$ & 0.18$^{+0.05}_{-0.04}$ & 0.23$^{+0.08}_{-0.06}$ &
0.14$^{+0.05}_{-0.03}$ \\
\enddata
\tablecomments{1$\sigma$ errors are shown. }
\tablenotetext{a}{Abundances are with respect to solar value \citep{ande89}.}
\end{deluxetable}

\end{document}